\documentclass[fleqn,usenatbib]{mnras}

\usepackage{newtxtext,newtxmath}

\usepackage[T1]{fontenc}

\DeclareRobustCommand{\VAN}[3]{#2}
\let\VANthebibliography\thebibliography
\def\thebibliography{\DeclareRobustCommand{\VAN}[3]{##3}\VANthebibliography}

\usepackage{graphicx}	
\usepackage{amsmath}	

\title[The excitation of NO by para-H$_2$ at low temperature]{Fine and hyperfine excitation of nitric oxide by collision with $para$-H$_{2}$
at low temperature}

\author[M. Ben Khalifa et al.]{
M. Ben Khalifa \thanks{E-mail: malek.benkhalifa@kuleuven.be}
and J. Loreau \thanks{E-mail: jerome.loreau@kuleuven.be}
\\
KU Leuven, Department of Chemistry, Celestijnenlaan 200F, B-3001 Leuven, Belgium.\\
}

\pubyear{2021}

\begin{document}
\label{firstpage}
\pagerange{\pageref{firstpage}--\pageref{lastpage}}
\maketitle

\begin{abstract}
Nitric oxide is an open-shell molecule abundantly detected in the interstellar medium. A precise modeling of its radiative and collisional processes 
opens the path to a precise estimate of its abundance. We
present here the first rate coefficients for fine and hyperfine (de-)excitation of NO by collisions with the most
ubiquitous collision partner in the interstellar medium, $para$-H$_2$ hydrogen molecules, using a recently developed accurate
interaction potential. We report quantum scattering calculations for transitions involving the first 74 fine levels and the
corresponding 442 hyperfine levels belonging to both $F_1$ and $F_2$ spin-orbit manifolds. To do so, we have calculated cross
sections by means of the quantum mechanical close-coupling approach up to 1000 cm$^{-1}$ of total energy and rate coefficients from 5 to 100 K.
Propensity rules are discussed and the new NO-H$_2$ rates are compared to those
available in the literature, based on scaled NO-He rates. Large differences are observed between the two sets
of rate coefficients, and this comparison shows that the new collision rates must be used in interpreting NO emission lines. We
also examined the effect of these new rates on the NO excitation in cold clouds by performing radiative transfer calculations
of the excitation and brightness temperatures for the two NO lines at 150.176 and 250.4368 GHz. This shows that the local
thermodynamic equilibrium is not fulfilled for this species for typical conditions. We expect the use of the rates presented in this
study to improve the constraints on the abundance of NO.
\end{abstract}



\section{Introduction}
The number of nitrogen-bearing molecules observed in astrophysical clouds is continuously growing. Their abundance,
stability as well as their synthesis still preoccupy various communities focusing on theoretical and experimental astrochemistry.
Among these nitrogen-bearing molecules, NO (nitric oxide, or nitrogen monoxide) is of particular interest 
as it is a ubiquitous radical molecule in astrophysical environments. It is formed from the reaction between N and the OH radical, and 
plays the role of a reaction intermediate in the formation of molecular nitrogen N$_2$. It is therefore an important species in the synthesis of nitrogen hydrides \citep{leGal2014interstellar,hily2010nitrogen,akyilmaz2007depletion,herbst1973formation}. In addition, it plays a major role in the formation of hydroxylamine (H$_3$NO), which is a key molecule in the amino acid formation pathway, and it is also believed to be of crucial importance for primitive life on Earth \citep{santana2017nitric}.\\

NO was first observed in the interstellar medium thanks to its fine and hyperfine observed lines by
\citet{liszt1978microwave}, and it has since been identified in photon dominated regions \citep{jansen1995millimeter}, circumstellar envelopes \citep{quintana2013detection,Velilla2015}, dark molecular clouds \citep{gerin1992abundance}, star-forming regions \citep{Blake1986,ziurys1991nitric,Tremblay2018}, protostellar shocks \citep{Codella2018}, in comet P/Halley
\citep{wallis1988some}, as well as extragalactic sources \citep{martin2003first}. 
The abundance of NO has been modelled in various environments, and found to be particularly high in circumstellar enveloppes with an abundance relative to H$_2$ exceeding $10^{-6}$ \citep{Velilla2015}. The interstellar chemistry of NO has been well studied ( see e.g. \cite{leGal2014interstellar}). NO is formed through the neutral-neutral barrierless reaction between hydroxyl radicals and atomic nitrogen. It can then react with N to form N$_2$, or with C to form CN. As such, an accurate knowledge of its abundance is crucial to understand more complex nitrogen-bearing species.

In most astrophysical environments, local thermodynamic equilibrium (LTE) conditions are not fulfilled, meaning that the rotational levels only become thermalized at densities higher than those occurring in the interstellar medium. The accurate interpretation of emission spectra, which allows one to go from emission line intensities to column densities or relative abundances, therefore require a radiative transfer model that takes into account the collisional excitation of molecules by He atoms and H$_2$ molecules by means of temperature-dependent state-to-state rate coefficients.

In previous studies, \citet{Klos2008} and \citet{lique2009importance} investigated the collisional excitation of NO by He atoms. The rate coefficients computed in these studies have been used in non-LTE modelling of NO in various environments (e.g., \citet{Velilla2015}). From a computational viewpoint, collisions with He are much simpler to treat than collisions with H$_2$. Helium is thus often used as a proxy for H$_2$ in the ground rotational level, as \textit{para}-H$_2$ has spherical symmetry and 2 valence electrons and scaled molecule-He rate coefficients are employed instead of molecule-H$_2$ rates. 

In the present work, we aim to compute the rotational (de)-excitation of NO by collision with H$_2$ using a recent four-dimensional potential energy surface (PES) obtained by \citet{klos2017interaction}. Based on this PES, we conduct quantum dynamics calculations  (section ~\ref{section3}) to yield collision cross sections in the close coupling formalism, and derive fine and hyperfine rate coefficients. Finally, we analyze the impact of our new set of rate coefficients in section ~\ref{section5} by performing an application with a radiative transfer calculation for typical interstellar conditions.

\section{Spectroscopy of NO}
Nitric oxide is an open shell molecule with a ground $X^2\Pi$ electronic state that is divided into two spin-orbit manifolds traditionally labeled $F_1$ for the set of lower levels and $F_2$ for the upper levels. The degree of splitting is governed by the spin-orbit constant A$_{\textrm{so}}$. In Hund's case (a), such type of molecule is characterized by a positive spin-orbit constant (here, $A_{\textrm{so}}$=123.1393 cm$^{-1}$). Therefore, the lower and upper component are denoted by $^2{\Pi}_{1/2}$ and $^2{\Pi}_{3/2}$
respectively \citep{herzberg1950spectra}. These correspond to the parallel and opposite values, $\vert\Omega\vert$,
of molecule-fixed projections $\Sigma$ and $\Lambda$ of the electron spin $S$ and the electronic orbital angular momentum $L$, which is equal to
|$\Omega$|=1/2 and $\vert\Omega\vert=\vert\Sigma+\Lambda\vert=3/2$.
The spin-orbit level of rotation is further divided into two closely spaced pairs, of opposite parity $p$ (+) and (-), denoted $e$ and $f$ respectively, called  the $\Lambda$-doubling. For a doublet multiplicity state, the total parity is $+(-1)^{j-1/2}$for the $e$-labeled states and $-(-1)^{j-1/2}$ for the $f$-labeled states
\citep{brown1975labeling}. In this paper, we denote the rotational levels of NO by the total angular momentum $j$ of NO and the fine-structure manifold $F_i$($i=1, 2$),

Furthermore, the nitrogen atom ($^{14}$N) has a non-zero nuclear spin ($I$=1), which generates an hyperfine splitting.  
In this case, the rotational states of each $\Lambda$-doublets level are split into 
3 hyperfine levels (except for the $j =1/2$ level which is divided into 2 levels), denoted by a quantum number
$F$ varying between |$I-j$| and $I+j$
as well as the parity $p$.
\section{SCATTERING CALCULATIONS}\label{section3}
\subsection{Methods}
In this work, we focus on the quantum dynamical treatment of an open-shell molecule impacted by a diatomic molecule in a $^1\Sigma^+$ electronic state,
H$_2$, with the purpose of determining cross sections and rate coefficients between the lowest fine and hyperfine structure levels, following the
formalism presented by \citet{offer1990rotational}, \citet{offer1994rotationally}, \citet{groenenboom2009bound} and 
\citet{schewe2015rotationally}.\\
The hyperfine splitting of energy levels due to the coupling between the nuclear spin and molecular rotation is extremely weak compared to the energy
of rotation and the energy of collision, so it is assumed that the hyperfine levels of NO are almost degenerate.
The state-to-state hyperfine cross sections can be then computed from the $T$-matrix of the fine-structure using a recoupling approach according to the expression \citep{offer1994rotationally}:
\begin{multline}
 \sigma_{j'F'_ip'F'j'_2,jF_ipFj_2}=\frac{\pi}{k_i^2(2F+1)(2j_2+1)}\sum_{J_{T}}(2J_{T}+1) \\
 \sum_{ll'j_Rj'_R} |T^{J_T}_{j'F'_ip'F'j'_2j'_Rl',jF_ipFj_2j_Rl}l|^2 $ $ $ $ $ $ $ $ $ $ $ $ $ $ $ $ $ $ $ $ $ $ $ $ $ $ $ $ $ $ $ $ $ $ $ $ 
 $ $
\end{multline}
where $T^{J_T}$ denotes the nuclear-spin-free $T$-matrix elements, for a total angular momentum of the collision system
$J_{T}$ ($J_{T}=J+I$; $J=j+l$) and $I$ is the nuclear spin of the proton. The total angular momentum of the NO radical, including nuclear spin,
is denoted by $F=j+I$, and the rotational angular momentum $j_2$ of the H$_2$ collision partner equals 0 for \textit{para-}H$_{2}$.
The $T$-matrix elements are given by
\begin{multline}
T^{J_T}_{j'F'_ip'F'j'_2j'_Rl',jF_ipFj_2j_Rl}=\sum_{Jj'_{12}j_{12}}(-1)^{j_R+j'_R+l+l'+j_2+j'_2} \\
 \texttimes ([F][F'][j_{12}][j'_{12}][j_R][j'_R])^{1/2}(2J+1)
\begin{pmatrix}
 j & j_2 & j_{12} \\
 l & J & j_R 
\end{pmatrix} 
\begin{pmatrix}
j' & j'_2 & j'_{12} \\
 l' & J & j'_R 
\end{pmatrix} \\
\begin{pmatrix}
 j_R & j & J \\
 I & J_T & F 
\end{pmatrix} 
\begin{pmatrix}
 j'_R & j' & J \\
 I & J_T & F' 
\end{pmatrix} 
T^J_{j'F'_ip'j'_2j'_{12}l',jF_ipj_2j_{12}l}
\end{multline}\\
where $J_R$, $j_{12}$ and $l$ are quantum numbers which denote the vector sum of $j_2+l$ and $j+j_2$, and the orbital angular momentum of the collision system respectively, and $[x]=2x+1$.

In order to compute the state-to-state inelastic cross sections, the spectroscopy of NO has to be reproduced with a good precision.
To do so, we used the experimental spectral parameters for NO($X^2\Pi$, $v$=0) \citep{varberg1999far}:  rotational constant $B_e$=1.69611 cm$^{-1}$, spin-orbit coupling constant $A_{\textrm{so}}$=123.1393 cm$^{-1}$, $\Lambda$-doubling constants: $p_{\Lambda}=0.01172$ cm$^{-1}$ and $q_{\Lambda}=0.00067$ cm$^{-1}$.
For H$_2$, the rotational constant is given by $B_0=59.322$ cm$^{-1}$ \citep{huber2013molecular}.\\
The scattering calculations for NO-H$_2$ collisions were carried out using the accurate PES developed by \citet{klos2017interaction}, performed within the explicitly correlated coupled-cluster RCCSD(T)-F12a formalism. The accuracy of this PES has been recently established by state-of-the-art molecular beam experiments on NO-H$_2$ and NO-D$_2$ inelastic collisions \citep{Vogels2018,Tang2020,Quan2020}.

Integral inelastic cross sections were computed using a full close coupling approach \citep{alexander1985quantum} implemented in the HIBRIDON package. The cross sections are calculated on a grid of total energies (E$_{\textrm{tot}}$) up to 1000 cm$^{-1}$
with a small energy step $dE$ at low energy in order to accurately describe the resonances in the cross sections.  
Therefore, the energy step is dependent on the energy range. The ranges, denoted as ($E_{\textrm{min}}$,$E_{\textrm{max}}$,$dE$) where $E_{\textrm{min}}$ and $E_{\textrm{max}}$ are the lower and upper limits of energies, were constructed as follows (all in cm$^{-1}$):  
(0, 100, 0.1), (100, 200, 0.2), (200, 300, 0.5), (300, 500, 1.0), (500, 700, 5.0), (700, 1000, 10.0).\\
In order to solve the coupled equations, we used the hybrid propagator of \citet{alexander1987stable}, with an integration starting from a distance of 3.5 bohr up to 200 bohr. 
The precision of the integral cross sections with respect to the integration parameters and the number of partial waves was tested until reaching 1\% of convergence criteria. 
For NO, we chose a large rotational quantum number including all levels up to $j_1$=27.5 in order to converge the cross sections up to $j_1$=22.5 for the $F_1$ manifold and
up to $j_1$=21.5 for the $F_2$ manifold. For \textit{para}-H$_2$, the rotational basis is restricted to $j_2$=0. We carried out convergence tests at various energies to explore the validity of this approximation, as illustrated in Table~\ref{test}. At low energy, increasing the rotational basis of H$_2$ molecule by including the  $j_2$=2 state has a negligible impact on the cross sections while the CPU time and disk occupancy are five times larger. At high energy and for transitions involving highly excited NO levels, the difference can reach 10\%. Neglecting the $j_2=2$ level of H$_2$ can thus be seen as the main source of uncertainty of the present calculations.
Converged cross sections required to sum over partial waves with a total angular momentum up to $J$=35.5 for $E_{\textrm{tot}}$
around 50 cm$^{-1}$, 45.5 for $E_{\textrm{tot}}$ around 100 cm$^{-1}$, up to 70.5 for $E_{\textrm{tot}}=500$ cm$^{-1}$ and 85.5 for 1000 cm$^{-1}$, the highest energy considered in this work.
We computed the resolved state-to-state hyperfine cross sections for transitions 
between the 74 fine levels of NO and the corresponding 442 hyperfine levels belonging to both $^2\Pi_{1/2}$
and $^{2}\Pi_{3/2}$ spin-orbit manifolds. 

\begin{table}
\centering
\caption{Comparison between cross sections (in ~\AA$^2$) for the excitation of NO by \textit{para}-H$_2$($j_2$=0) calculated with a rotational basis set including only the $j_2=0$ rotational state of H$_2$, or the $j_2=0$ and $j_2=2$ states, for transitions within the $F_1$ manifold, for total energies $E_{\textrm{tot}}$=100, 500 and 800 cm$^{-1}$.}
\label{test}
\begin{tabular}{ccccc}
\hline
\hline
Energy (cm$^{-1})$ & $j_{2\max}$ & 0.5$f \rightarrow$ 0.5$e$ & 1.5$f \rightarrow$1.5$e$ & 2.5$e \rightarrow$ 1.5$f$ \\
\hline
$E=100$ & 0 & 0.783 & 0.485 & 3.100  \\
      & 	  2 & 0.734 & 0.442 & 3.191  \\
\hline
$E=500$ & 0 & 0.505 & 0.580 & 2.552  \\
      & 2 & 0.487 & 0.5436 & 2.562  \\
\hline
$E=800$ & 0 & 0.450  & 5.243 & 1.047  \\
      & 2 & 0.403 & 4.965 & 1.062  \\

\hline
\hline
\end{tabular}
\end{table}

\subsection{Cross sections}
\begin{figure*}
\centering
{\label{a}\includegraphics[width=.46\linewidth]{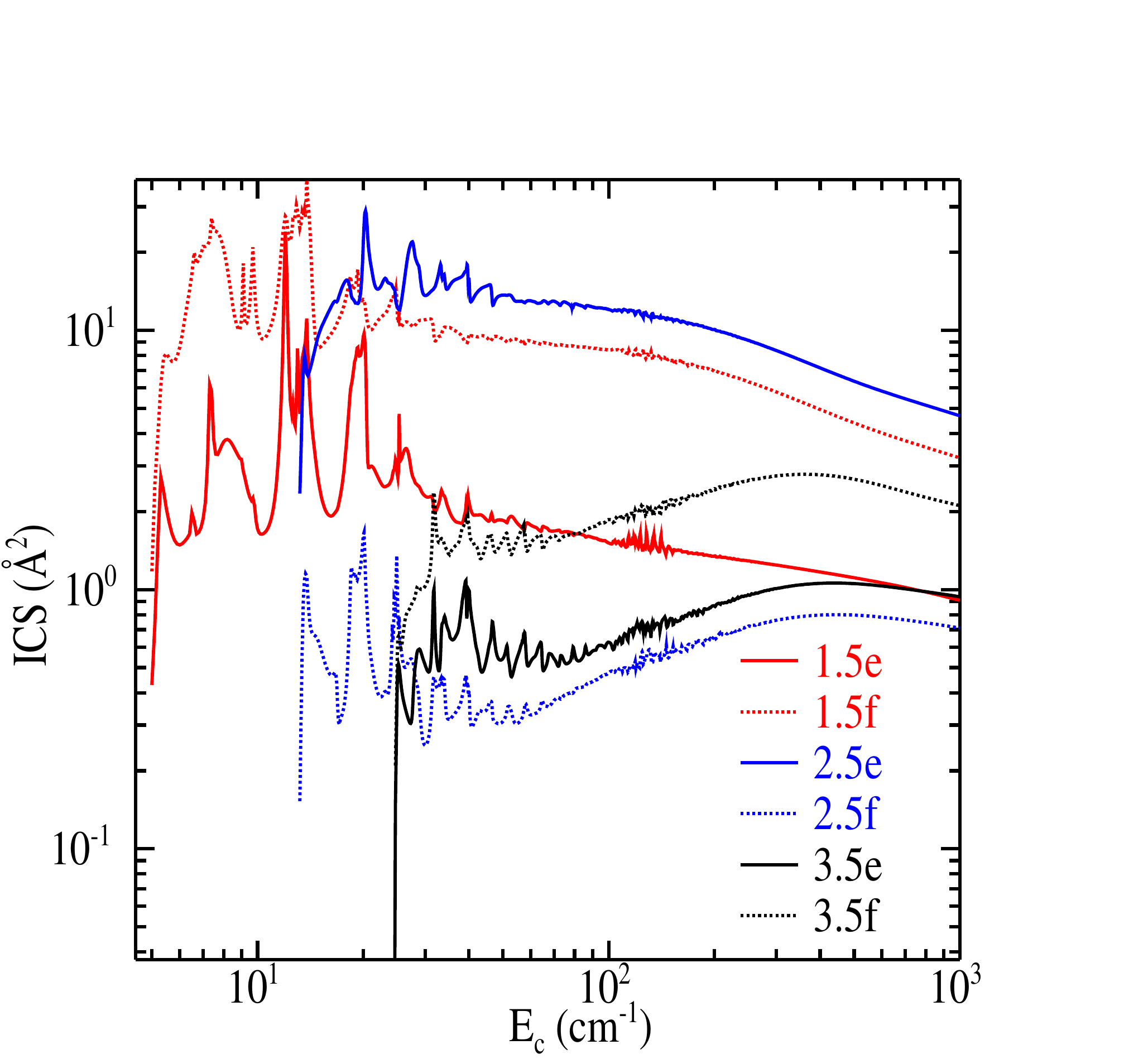}}\hfill
{\label{b}\includegraphics[width=.47\linewidth]{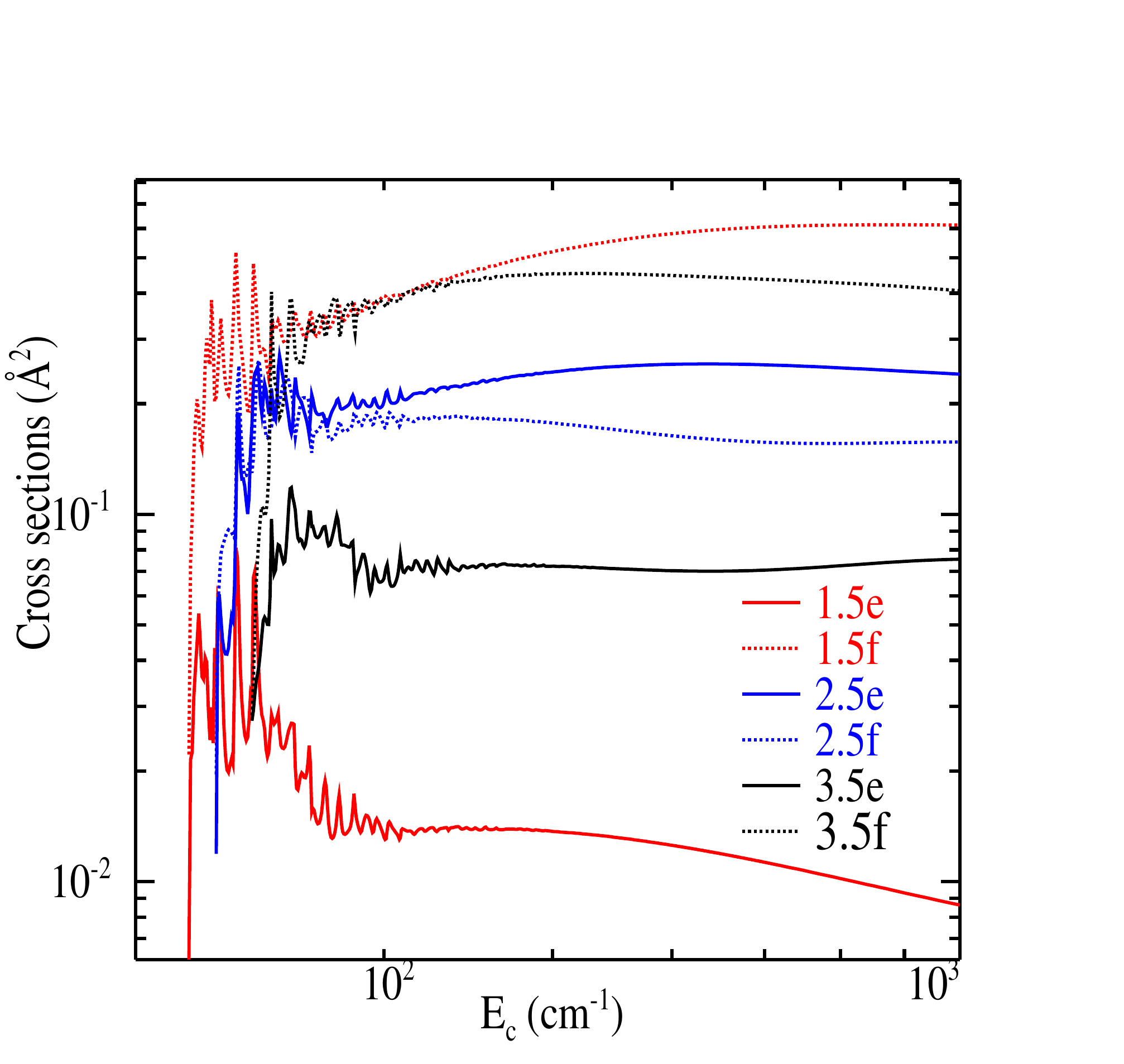}}\par 
\caption{Collision energy dependence of cross sections for transitions out of the lowest ($j= 0.5,e$) fine structure level of the $F_1$ spin-orbit
manifold to several rotational levels ($j,e/f$) within both the $F_1$(left panel) and $F_2$(right panel) manifold.}\label{XCS}
\label{fig:level} 
\end{figure*}

Figure~\ref{XCS} highlights the kinetic energy dependence of the excitation cross section for transitions from the ground fine structure level
($j_1= 0.5,e$) of the lower $^2\Pi_{1/2}$ spin-orbit manifold to excited states of NO ($j_1',e/f$) in both $^2\Pi_{1/2}$ and $^2\Pi_{3/2}$ manifolds.
As may be seen, these curves present a noticeable oscillatory structure at low energy ($E_{\textrm{coll}}$ $\leq$ 100 cm$^{-1}$), which are expected given the well depth of the PES ($D_e \approx 80.4$ cm$^{-1}$ \citep{klos2017interaction}).
These oscillations present shape and/or Feshbach resonances character. These low energy resonances were recently investigated experimentally with a crossed-molecular-beam technique by \citet{Vogels2018} in the case of the $0.5f \rightarrow 1.5e$ transition in the $^2\Pi_{1/2}$ manifold. The experimental findings were supported by scattering calculations on the same PES as the one employed in the present paper, and our results for this particular transition are in near perfect agreement.

From figure~\ref{XCS}, it is obvious that transitions that conserve the spin-orbit are about one order of magnitude larger than transitions accompanied by a spin-orbit change. However, the cross sections for transitions between the $F_1$ and $F_2$ manifolds can never be considered negligible.
Furthermore, we note that inelastic cross sections for transitions that conserve spin-orbit ($F_1 \leftrightarrow F_1$ and $F_2 \leftrightarrow F_2$) 
present a strong propensity in favour of the parity conservation ($e \leftrightarrow e$ or $f \leftrightarrow f$) and with $\Delta j_1=2$, and in favor of transitions
with odd $\Delta j_1$ when a change of parity occurs ($e \leftrightarrow f$).
The dominance of transitions with even $\Delta j$ is related to the interference effect, due to the shape of the PES which is symmetrical with
respect to $\theta_1=90^{\circ}$ \citep{mccurdy1977interference,klos2017interaction}.
By contrast, for spin-orbit changing transitions ($F_1 \leftrightarrow F_2$), the cross sections corresponding to even/odd $\Delta j_1$ transitions or to parity-changing or parity-conserving transitions have magnitudes that do not follow simple propensity rules. \\

\subsection{Fine and hyperfine rate coefficients}\label{section4}
Rate coefficients were calculated by averaging cross sections for initial and final states $i$ and $f$, $\sigma_{i \rightarrow f} (E_c$),  over the Maxwell-Boltzmann distribution of collision energies as
expressed :
\begin{equation}
 k_{i \rightarrow f}(T)=\biggl(\frac{8}{\pi\mu\beta}\biggl)^{\frac{1}{2}}\beta^2\int_0^{\infty} E_c \sigma_{i \rightarrow f}(E_c)e^{-\beta E_c} dE_c
\end{equation}
Where $\beta$=$\frac{1}{k_BT}$ and $k_B$, $T$ and $\mu$ denote the Boltzmann constant, the kinetic temperature and the NO-H$_2$ reduced mass, 
respectively.\\
As cross sections were computed for total energies up to 1000 cm$^{-1}$, we were able to compute the rate coefficients for transitions
involving the first 442 hyperfine levels 
for temperatures ranging between 5~K and 100~K. Tables of rate coefficients are available as supplementary material. 

An overview of the fine structure rate coefficients for spin-orbit conserving and spin-orbit changing transitions in NO-H$_2$ collisions
is presented in Figure ~\ref{ref:fine}.
As a first observation, rate coefficients for spin-orbit changing transitions are in general much smaller than those for spin-orbit conserving transitions, corresponding to the behavior observed for the cross sections. Similarly, among spin-orbit conserving transitions, those that conserve parity strongly dominate for even values of $\Delta j_1$.
Moreover, a conservation of the total parity is found for transitions accompanied by a change in spin-orbit. We can see from figure ~\ref{ref:fine}
that rate coefficients are larger for parity breaking transitions and a propensity to populate final levels $(e/f)$ when starting from initial levels $(f/e)$. 

There are no rate coefficients available in the literature for NO-H$_2$, although \citet{Klos2008} computed fine structure rate coefficients for NO-He. Since \textit{para}-H$_2$($j_2$=0) is spherical symmetric and isoelectronic to He, it is often assumed that the excitation of molecules in collisions with \textit{para}-H$_2$($j_2$=0) can be modeled by using collisions with He as a proxy, once the proper scaling is used to take into account the difference in reduced mass (factor of 1.39). 

\begin{figure*}
\centering
{\label{b}\includegraphics[width=.5\linewidth]{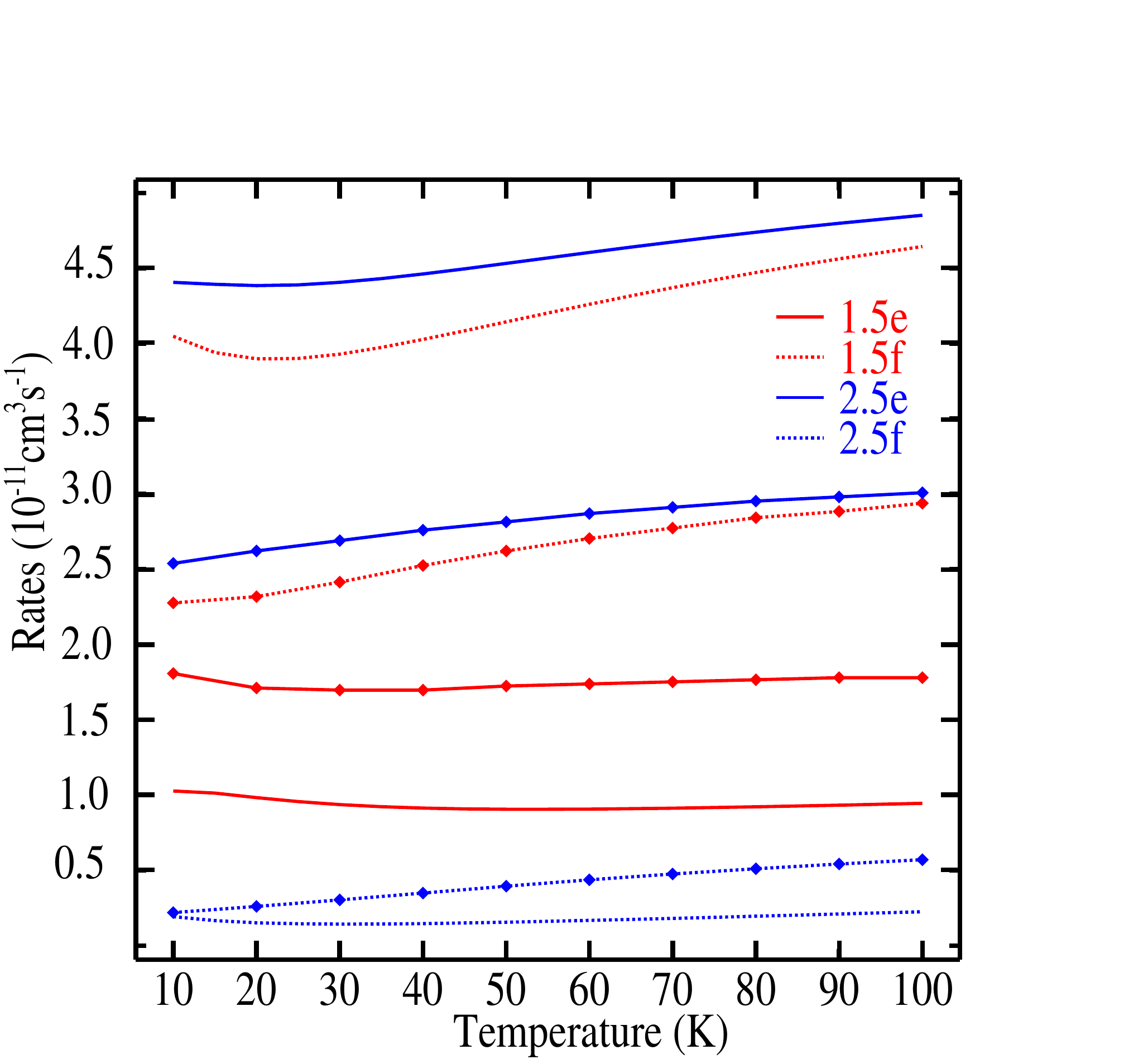}}
\caption{Comparison of the fine structure resolved de-excitation rate coefficients in collisions of NO with H$_2$($j_2$=0) (lines) and He (diamonds) collisions, out 
of various $(j',e/f)$ levels within the $F_1$ spin-orbit manifolds to the ground state $^2\Pi_{1/2}(j=0.5e)$ level.}\label{ref:fine}
\label{fig:compHefine} 
\end{figure*}

In Fig. \ref{fig:compHefine} we present a comparison of the fine structure resolved de-excitation rate coefficients of NO-H$_2$ and scaled NO-He for transitions between various $(j',e/f)$ levels to the $^2\Pi_{1/2}(j=0.5e)$ level. The rates for NO-H$_2$ are usually larger than the scaled NO-He rates, although there are exceptions. From this figure it is already clear that the scaling of collisional rates with He cannot be used to model collisions with H$_2$ for the present system. Several factors can contribute to explain this difference. Firstly, while the PESs for NO-He and NO-H$_2(j_2=0)$ present qualitatively similar anisotropies, the depth of the PES is almost three times larger in the case of H$_2$
($D_e \approx 29.2$ cm$^{-1}$ for NO-He \citep{kl2000ab} and $D_e \approx 80.4$ cm$^{-1}$ for NO-H$_2$
\citep{klos2017interaction}). The potential  is also more repulsive in the short-range, which is expected to strongly impact the magnitude of the cross sections and rate coefficients.

\begin{figure*}
\centering
{\label{a}\includegraphics[width=.46\linewidth]{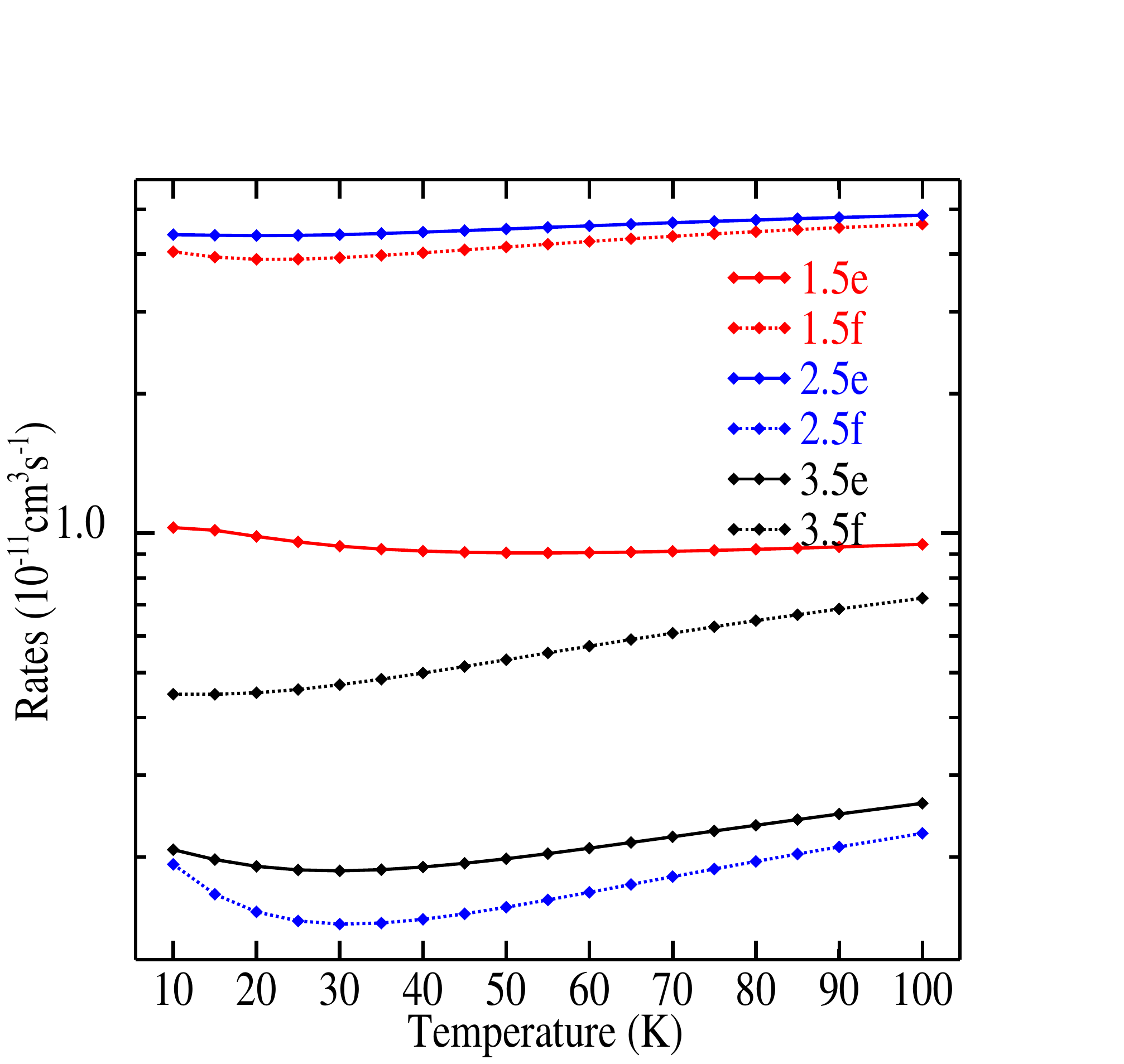}}\hfill
{\label{b}\includegraphics[width=.47\linewidth]{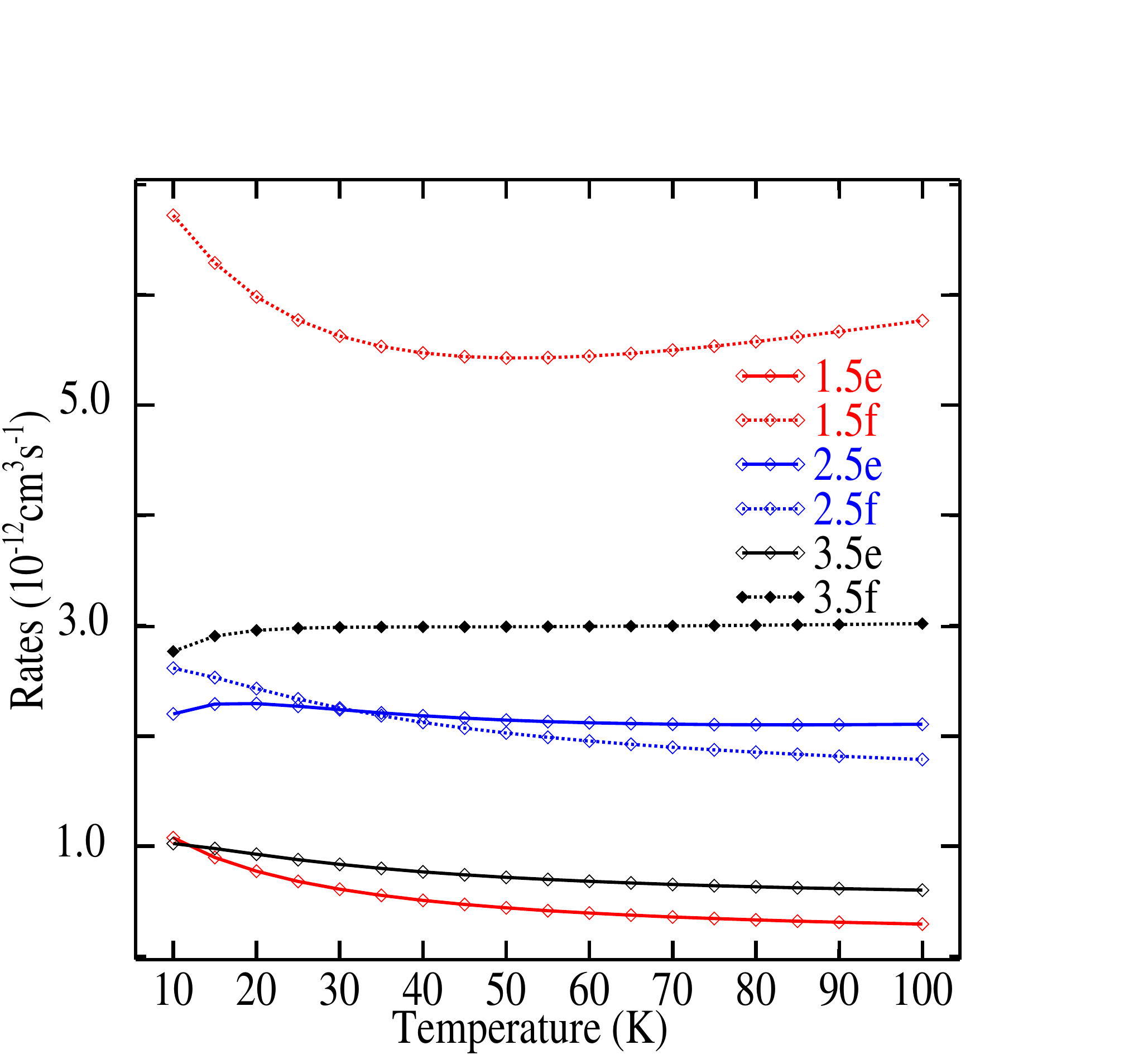}}\par 
\caption{Temperature dependence of the fine structure resolved NO-H$_2$($j_2$=0) collisional de-excitation rate coefficients out 
of various $(j',e/f)$ levels within both the $F_1$ (left panel) and $F_2$ (right panel) spin-orbit manifolds to the ground state $^2\Pi_{1/2}(j=0.5e)$ level.}\label{ref:fine}
\label{fig:level} 
\end{figure*}

\begin{figure*}
\centering
{\label{a}\includegraphics[width=.46\linewidth]{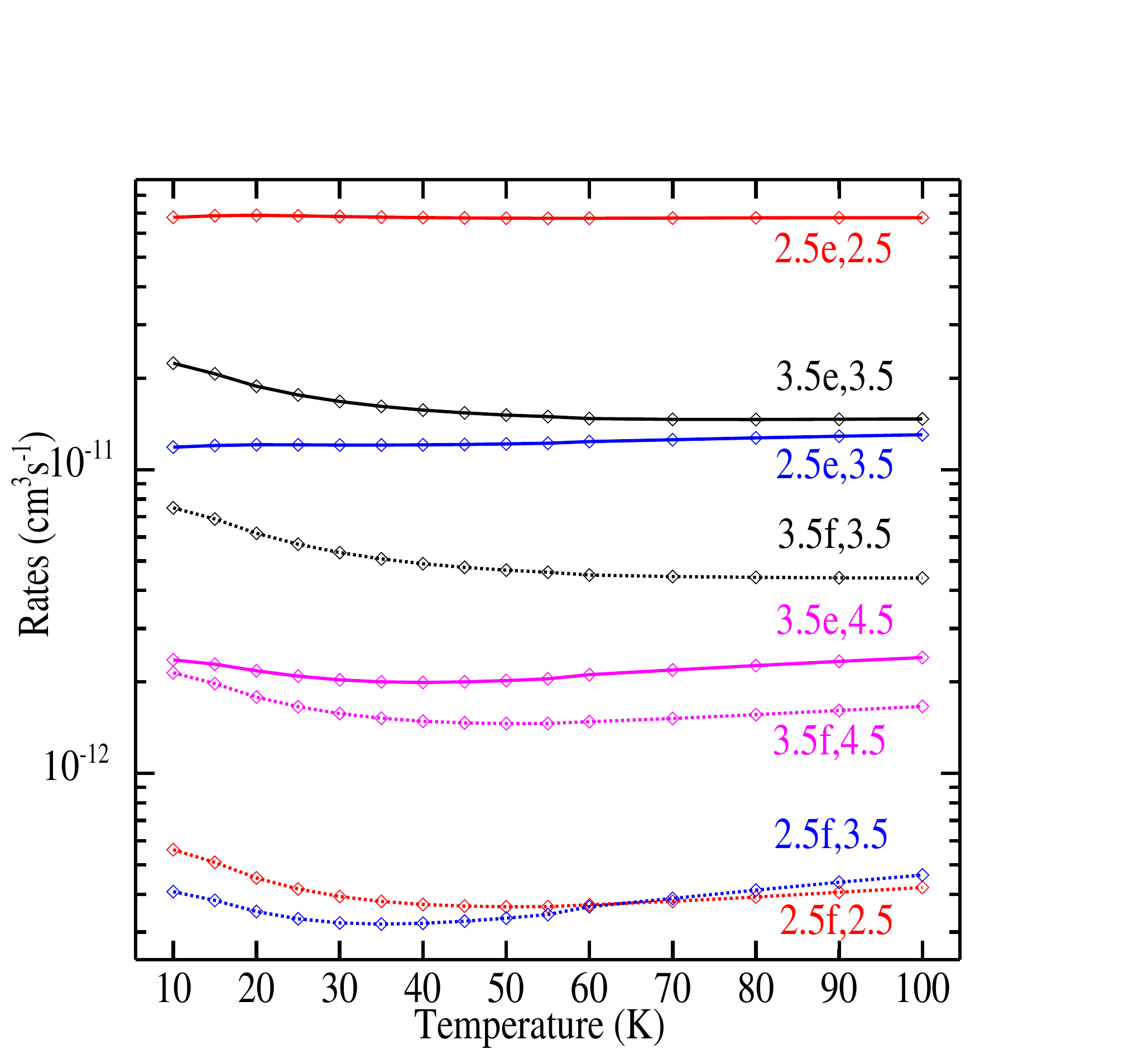}}\hfill
{\label{b}\includegraphics[width=.47\linewidth]{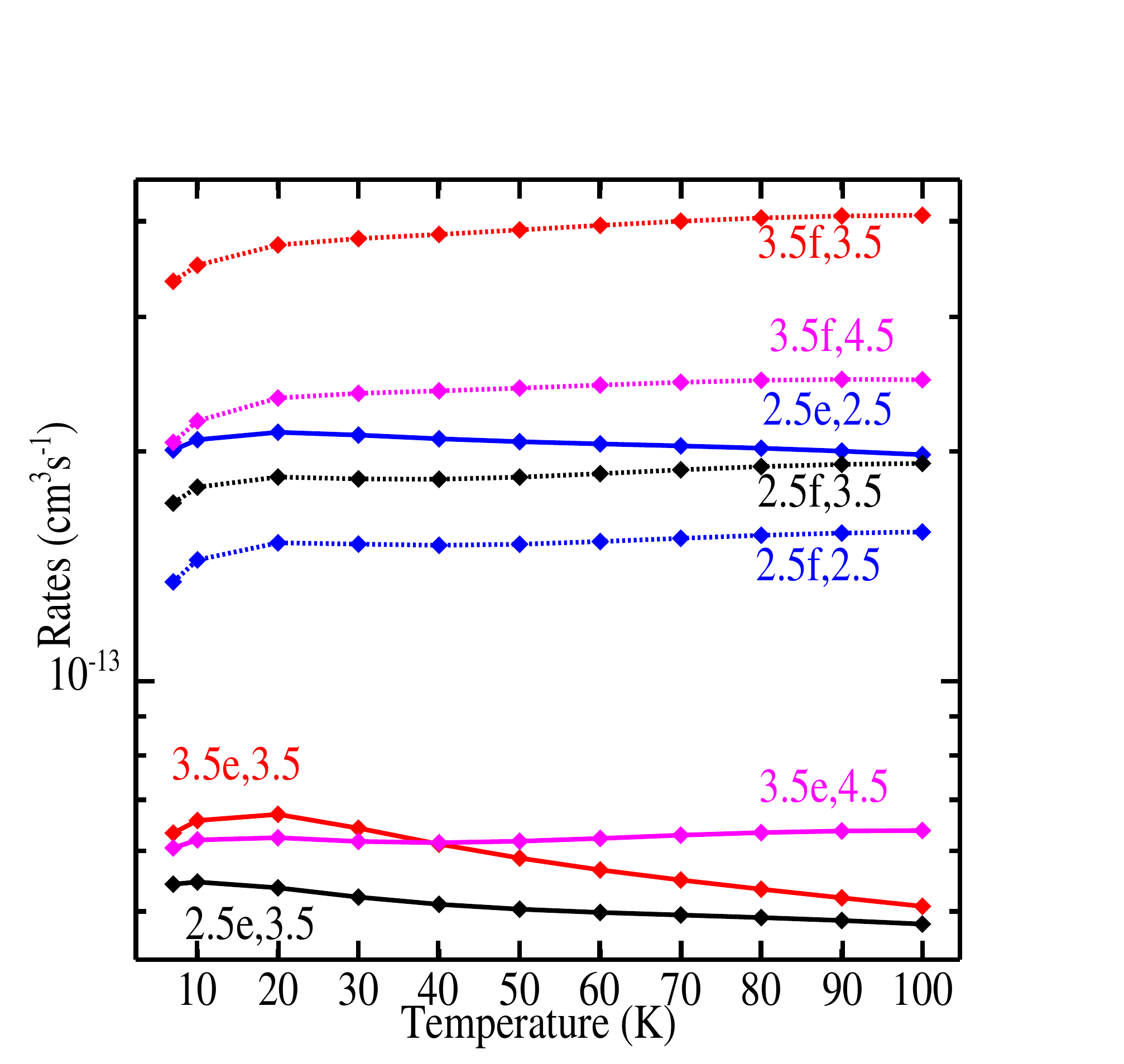}}\par 
\caption{Rate coefficients for collision of NO with H$_2$ for transitions out of the $j_1=4.5,e,F=4.5$ hyperfine level in the $\Omega=1/2$
(left panel) and $\Omega=3/2$ manifolds (right panel). The parity-conserving transitions are shown in solid line and the final levels $(j', p',F')$ are indicated.}
\label{fig:Hyperfine} 
\end{figure*}

Figure~\ref{fig:Hyperfine} presents a representative example of the temperature variation of the hyperfine collisional rates, for transitions out of $j_1=4.5e,F=4.5$ hyperfine level within the $F_1$ and $F_2$ spin-orbit manifolds. As one would anticipate, the hyperfine rates for parity-conserving transitions present the same propensity rules as the fine structure rates, with rate coefficients for spin-orbit transitions being larger than those with a change in spin orbit by up to one order of magnitude.
In addition, these transitions clearly show the usual hyperfine propensity rules, in fact, transitions with $\Delta j=\Delta F$ are always dominant. This behavior has already been observed in several systems such as CCN-He, C$_6$H-He, or CN-H$_2$ \citep{chefai2020fine, walker2018fine, kalugina2012hyperfine}.

\begin{figure*}
\centering
{\label{b}\includegraphics[width=.47\linewidth]{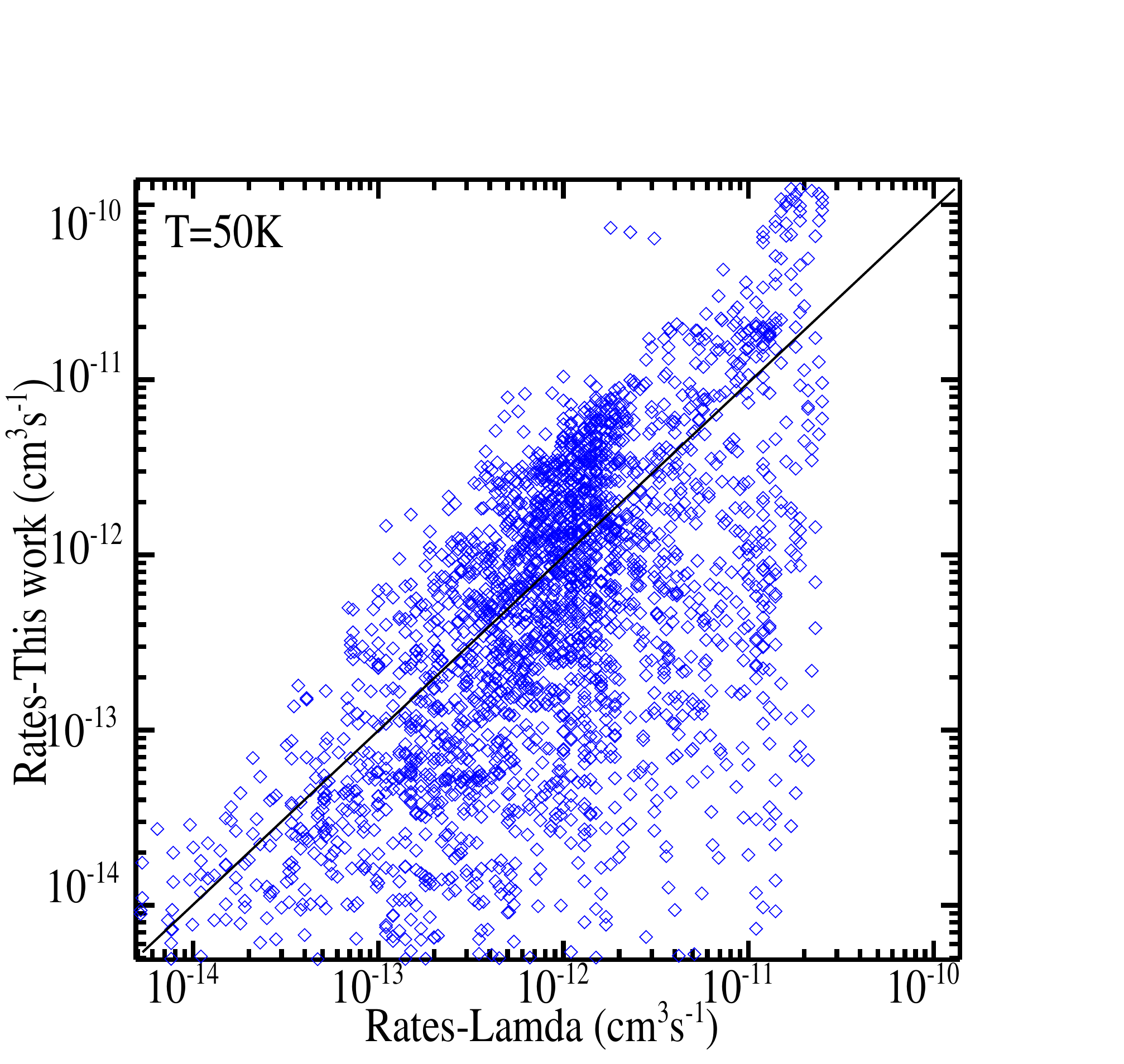}}
\caption{Comparison of hyperfine rates between present work and the published LAMDA rates based on scaled NO-He rate coefficients.}
\label{fig:comp} 
\end{figure*}

We can again compare our results for NO-H$_2$ with those for NO-He. Quenching rates calculated at a temperature of 50~K are presented in figure~\ref{fig:comp} for transitions among the first 442 NO hyperfine levels and compared with those of \citet{lique2009importance}.
It is noticeable that the rate coefficients differ strongly, with differences of up to three orders of magnitude.
In addition to the differences in PESs discussed above, another explanation for the dramatic difference between hyperfine rate coefficients of these systems is that the hyperfine rates for NO-He were obtained with a statistical approximation that assumes the hyperfine  rate coefficients to be proportional to the degeneracy of the final hyperfine level, which does not appear be valid for NO-H$_2$. \\

The present cross sections were also validated by comparison with the measurements and close-coupling calculations performed by \citet{Vogels2018} with a crossed-beam apparatus.

The large difference between the rate coefficients for NO-H$_2$ compared to the scaled NO-He rates therefore calls for a re-examination of their impact on the abundance of NO in various astronomical environments.

\section{ASTROPHYSICAL APPLICATION}\label{section5}

In this section, we apply the hyperfine excitation rate coefficients to estimate their effect on the abundance of NO in the ISM by carrying out non-local thermodynamic  equilibrium (non-LTE) radiative transfer computations.

The hyperfine lines of NO are resolved in several interstellar sources \citep{ziurys1991nitric} \citet{gerin1993abundance},
and the corresponding intensities can be employed as tracers of the physical conditions in this environment.
The \verb+RADEX+ code \citep{van2007computer} is employed to compute the intensities of molecular transitions assuming a uniform spherical geometry of a homogeneous interstellar medium. 
We take into consideration both radiative and collisional processes, while the optical depth impacts are modeled within an escape probability formalism approximation.\\
We started by calculating a collider critical density $n^*$(H$_2,T)$ using the following equation :\\
\begin{equation}
n^*(H_2,T)=\frac{A_{fi}}{\sum_{f \neq i}k_{fi}}
\end{equation}
where $A_{fi}$ and $k_{fi}$ are the Einstein spontaneous absorption coefficients and rates respectively. 
This was performed for transitions corresponding to emission in the 50-500 GHz interval, which is easily observable with ground-based radio telescopes. This includes in particular the often observed $^2\Pi_{1/2}$ (1.5$f$,2.5 - 0.5$e$,1.5) transition at 150.176 Ghz and the $^2\Pi_{1/2}$ (2.5$e$,3.5 - 1.5$f$,2.5) transition at 250.4368 Ghz. The critical densities (in cm$^{-3}$), calculated at $T = $10, 50 and 80 K, are presented in table ~\ref{nc}. 
We note that the critical density of these detected transitions revolves around 10$^3$-10$^4$ cm$^{-3}$. These results correspond to the typical density of the interstellar gas, accordingly, the local thermodynamic equilibrium is far from being reached, especially for transitions involving high rotational levels. The use of non-LTE simulations is therefore crucial in order to correctly model the NO emission spectra.

\begin{table}
\centering
\caption{NO critical densities $n_c^*$(cm$^{-3}$) at 10, 50 and 80 K for the observed transitions.}\label{nc}
\begin{tabular}{cccc}
\hline
\hline
Transition & $T=10$ K & $T=50$ K & $T=80$ K \\
\hline
1.5$f$,2.5 $\rightarrow$ 0.5$e$,1.5 & $0.200 \times 10^3$ & $0.220 \times 10^3$ &  $0.211 \times 10^3$ \\
2.5$e$,3.5 $\rightarrow$ 1.5$f$,2.5 & $0.977 \times 10^3$ & $0.113 \times 10^4$ &  $0.108 \times 10^4$ \\
\hline
\hline
\end{tabular}
\end{table}

\begin{figure*}
\centering
{\label{a}\includegraphics[width=1.\linewidth]{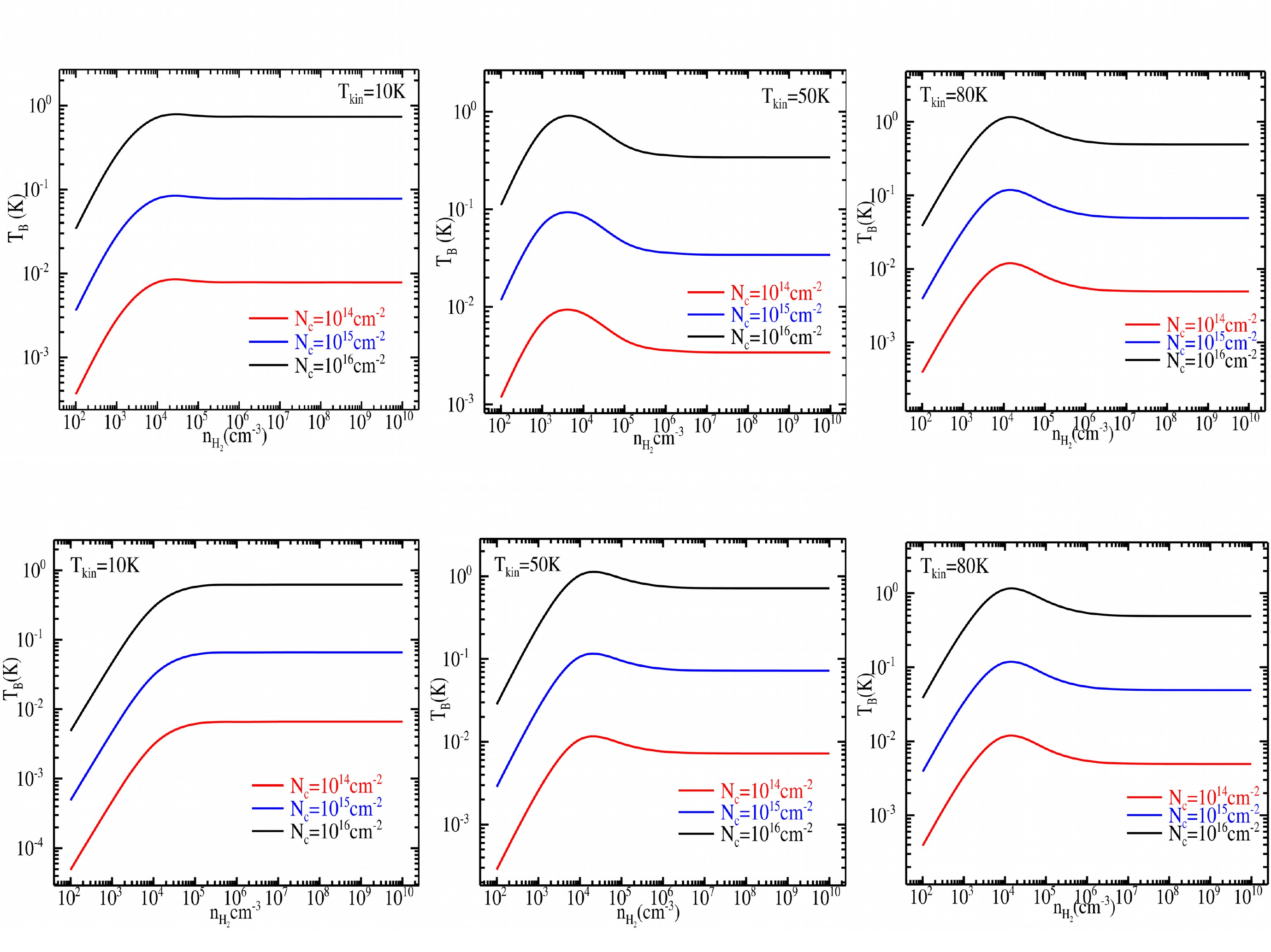}}\hfill
\caption{Brightness temperature for the two NO hyperfine transitions, in the 150.176 GHz (1.5$f$,2.5 $\rightarrow$ 0.5$e$,1.5)
(top panel) and 250.4368 GHz (2.5$e$,3.5 $\rightarrow$ 1.5$f$,2.5) (bottom panel) as a function of the H$_2$ density for three kinetic temperature
10~K, 50~K and 80~K.}
\label{fig:Brightness} 
\end{figure*}

In the radiative transfer simulation, we set the basic parameters as follows: a 2.7 K cosmic microwave background (CMB) as a background radiation field, and a line width of 8.0 km.s$^{-1}$. We varied the density of the molecular hydrogen between 10$^{2}$ and 10$^{10}$ cm$^{-3}$ and the column  density of NO in the 10$^{14}$-10$^{16}$ cm$^{-2}$ range, a choice that is based on the estimated column density of the molecule in the dark cloud L134N \citep{mcgonagle1990detection}.
The variation of the brightness temperature as a function of the H$_2$ density and for various NO column densities is shown in Figure~\ref{fig:Brightness}. For very low volume densities, we observe a linear dependence of T$_B$, before stabilizing at high densities in the 10$^6 - 10^8$ cm$^{-3}$ range. T$_B$ increases with increasing column density, with an asymptotic behaviour for densities larger than about 10$^6$ cm$^{-3}$ due to the small opacities of the lines ($\tau \leq$ 1).

\begin{figure*}
\centering
{\label{a}\includegraphics[width=1.\linewidth]{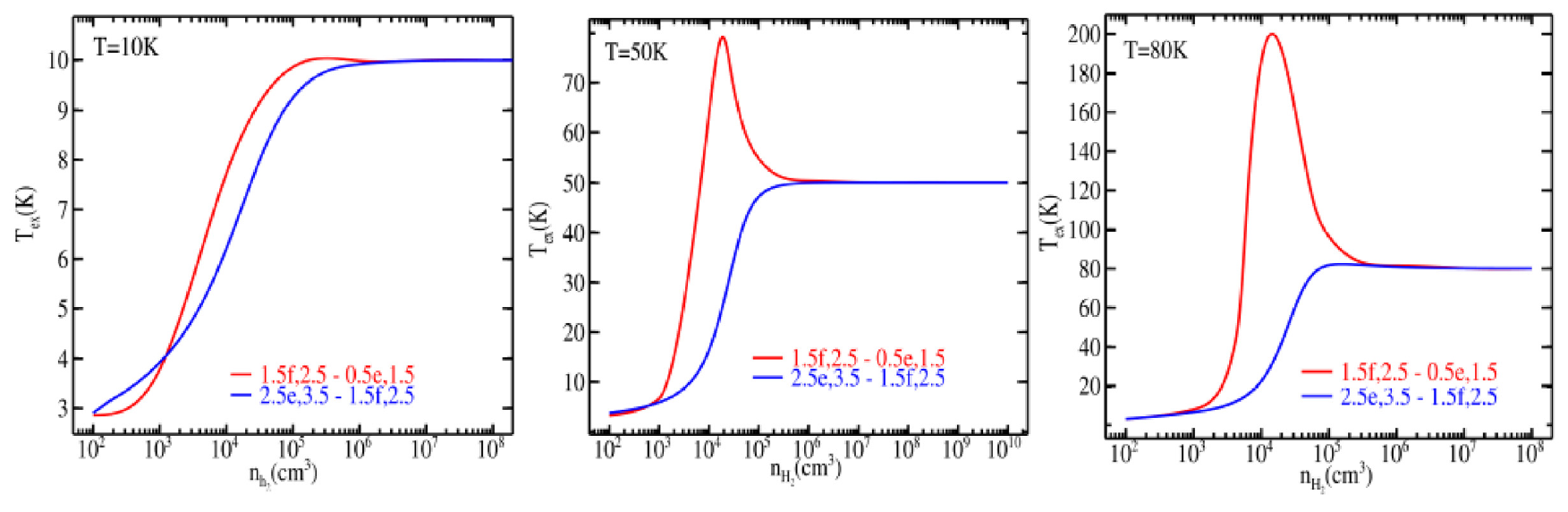}}\hfill
\caption{Excitation temperature for the two NO hyperfine transitions, in the 150.176 GHz (1.5$f$,2.5 $\rightarrow$ 0.5$e$,1.5)
and 250.4368 GHz (2.5$e$,3.5 $\rightarrow$ 1.5$f$,2.5) as a function of the H$_2$ density for three kinetic temperature
10~K, 50~K and 80~K and column density of 10$^{14}$ cm$^{-2}$.}
\label{fig:Excitation} 
\end{figure*}

Figure \ref{fig:Excitation} presents the excitation temperature as determined in our computations using a fixed NO column density of 10$^{14}$ cm$^{-2}$, which corresponds to the typical NO abundance in the dark cloud TMC-1 \citep{gerin1993abundance}. 
Computations for densities of 10$^{15}$ cm$^{-2}$ and 10$^{16}$ cm$^{-2}$ gave almost identical results, as the NO transitions are optically thin for the conditions considered here.
In addition, we note that the excitation temperature of the transitions in the 150.176 Ghz: $^2\Pi_{1/2}$ (1.5$f$,2.5 - 0.5$e$,1.5)
and 250.4368 Ghz: $^2\Pi_{1/2}$ (2.5$e$,3.5 - 1.5$f$,2.5) is equal to the value of the background radiation field (2.7 K ) at low
H$_2$ densities, and gradually increases to higher values as the collisional excitation process becomes more important.
At high volume densities, the excitation temperature tends asymptotically towards the kinetic temperature, at which point the LTE is reached and the populations of the rotational levels no longer depend on the density of H$_2$ and simply obey Boltzmann's law.
For the transitions at 150.176 Ghz and 250.4368 GHz, we can consider that the LTE is reached for densities above 10$^6$~cm$^{-3}$.
These values are greater than the typical density of many regions of the ISM (10$^3 \leq n(\textrm{H}_2) \leq 10^5$~cm$^{-3}$), which demonstrates that the NO transitions are not thermalized and that non-LTE models should be employed to analyse emission spectra.
We note in particular a suprathermal effect for the emission line (1.5$f$-2.5 - 0.5$e$-1.5) at kinetic temperatures $T_{k}=50$~K and $80$~K.
In these cases, the excitation temperature reaches a peak with excitation temperatures much higher than the kinetic temperature at densities around $10^4$ cm$^{-3}$. 
This behavior is common for the excitation of diatomic molecules and was also observed for NO-He \citep{lique2009importance}. 
For the two lines considered here, the excitation temperature is similar to that obtained for the excitation of NO by He by \citet{lique2009importance}. On the other hand, the brightness temperature is smaller for NO-H$_2$ than for NO-He for identical physical conditions.

\section{conclusion}
We have calculated a new set of fine- and hyperfine-resolved rate coefficients corresponding to the excitation of NO in collisions with \textit{para}-H$_2$($j_2$=0) using the quantum  close-coupling approach based on a very precise
4D-PES computed recently by \citet{klos2017interaction}. Collisional rates were obtained for transitions involving the lowest 442 hyperfine levels of rotation within of the two spin-orbit manifolds $^2\Pi_{1/2}$ and $^2\Pi_{3/2}$ for kinetic temperatures up to 100 K. \\
The new NO-H$_2$ rate coefficients show important differences with those available for NO-He that are currently used for astrophysical modeling. We therefore recommend the use of these new rates in non-LTE models of NO excitation.
We have also used the new rate coefficients in a simple radiative transfer model to assess their impact on the two transitions $^2\Pi_{1/2}$ (1.5$f$,2.5 - 0.5$e$,1.5) at 150.2 Ghz and $^2\Pi_{1/2}$ (2.5$e$,3.5 - 1.5$f$,2.5) at 250.4 Ghz. The critical densities for the NO levels of rotation were determined to be around 10$^5$ cm$^{-3}$, which demonstrates that a non-LTE analysis is important to model the NO emission spectra.

The full set of rate coefficients will be made available in the LAMDA \citep{schoier2005atomic} and BASECOL \citep{dubernet2013basecol2012} databases,
as well as supplementary material.

\section*{Acknowledgements}
The help of P. Dagdigian with the HIBRIDON program is gratefully acknowledged.
The scattering calculations presented in this work were performed on the VSC clusters (Flemish Supercomputer Center), funded by the Research Foundation-Flanders (FWO) and the Flemish Government, as well as on the Dirac cluster of KU Leuven.
J.L. acknowledges financial support from Internal Funds KU Leuven through grant 19-00313.

\section*{DATA AVAILABILITY}
The data underlying this article are available as supplementary material.





\bsp	
\label{lastpage}
\end{document}